\documentclass[pre,preprint,superscriptaddress,showpacs,preprintnumbers,eqsecnum,amsmath,amssymb]{revtex4}
\usepackage{graphicx}
\usepackage{dcolumn}
\usepackage{bm}
\usepackage{color}
\usepackage{epsfig}
\usepackage{psfrag}

\definecolor{darkgreen}{rgb}{0,0.6,0}

\newcommand{\comment}[1]{}
\input xy 
\xyoption{all}

\begin{document}

\title{
Wave Transport in One-Dimensional Disordered Systems   \\
with Finite-Size Scatterers 
}

\author{
Marlos D\'iaz \footnote{Deceased}
}
\affiliation{
Instituto de F\'{\i}sica, Universidad Nacional Aut\'{o}noma de M\'{e}xico,
Apartado Postal 20-364, M\'{e}xico, D.F.
}

\author{
Pier A. Mello
}
\affiliation{
Instituto de F\'{\i}sica, Universidad Nacional Aut\'{o}noma de M\'{e}xico,
Apartado Postal 20-364, M\'{e}xico, D.F.
}

\author{Miztli Y\'epez}
\affiliation{
Departamento de F\'isica, Universidad Aut\'onoma Metropolitana-Iztapalapa,
Apartado Postal 55-534, 09340, M\'{e}xico, D.F.
}

\author{
Steven Tomsovic
}
\affiliation
{
Department of Physics and Astronomy,
Washington State University,
Pullman, WA
}

\date{\today}

\begin{abstract}

We study the problem of wave transport in a one-dimensional disordered system, where the  scatterers of the chain are $n$ barriers and wells with statistically independent intensities and with a spatial extension $\l_c$ which may contain an arbitrary number 
$\delta/2\pi$ of wavelengths, where $\delta = k l_c$. 
We analyze the average Landauer resistance and transmission coefficient of the chain as a function of $n$ and the phase parameter $\delta$.  For weak scatterers, we find: 
i) a regime, to be called I, associated with an exponential behavior of the resistance with $n$,
ii) a regime, to be called II, for $\delta$ in the vicinity of $\pi$, where the system is almost transparent and less localized, and
iii) right in the middle of regime II, for $\delta$ very close to $\pi$, the formation of a band gap, which becomes ever more conspicuous as $n$ increases.
In regime II, both the average Landauer resistance and the transmission coefficient show an oscillatory behavior with $n$ and $\delta$.  
These characteristics of the system are found analytically,
some of them exactly and some others approximately.
The agreement between theory and simulations is excellent,
which suggests a strong motivation for the experimental study of these systems.
We also present a qualitative discussion of the results.
\end{abstract}


\maketitle

\section{Introduction}
\label{intro}

The problem of wave transport in disordered systems has been extensively studied in the literature, 
both for uncorrelated disorder
(see, e.g., 
Ref.~\cite{anderson58,lifshitz,sheng,altshuler,beenakker,mott,mello-kumar,froufe_et_al_2007} 
and references therein), 
as well as for the case in which the disordered potential shows correlations
\cite{lifshitz,dunlap,bovier92,flores-hilke,evangelou93,izrailev,moura98,titov05}.



Common features of the problems investigated by our group in 
Refs. \cite{mello-kumar,froufe_et_al_2007} are that 
i) uncorrelated disordered is contemplated, and 
ii) the size of the individual scatterers that compose the disordered system is the smallest one occurring in the problem: 
in particular, it is much smaller than the wavelength of the wave sent along the waveguide, and is thus of no physical relevance.
In these models, each individual potential, statistically independent from the others, is modeled by a delta function, and the distance between successive scatterers is subsequently
taken to be very small, which allows considering the so-called dense weak-scattering limit (DWSL), an important ingredient in the analysis carried out in those references.
Various quantities of physical interest were investigated within this framework, like the conductance, its fluctuations, and the individual transmission coefficients of the disordered system. 
A particularly attractive property that was found is the insensitivity of the results to details of the individual-scatterer statistical distribution, expressed in the form of a central-limit theorem.

In the present paper we build on previous work \cite{marlos_et_al} to study the simplest extension of the problems contemplated in Refs. \cite{mello-kumar,froufe_et_al_2007}:
the problem of wave transport in 1D disordered systems, in which the various scatterers have a finite size.
Specifically, we consider a succession of $n$ barriers and wells, to be referred to, generically, as steps, having a finite width.
The potential under study is shown schematically in Fig. \ref{rand_steps} below.
It contains $n$ steps, assumed to be weak compared with the energy $E$.
The steps are characterized by:  

i) A fixed width $l_c$ which may fit an arbitrary number of wavelengths 
$\delta/2 \pi$, where the parameter $\delta = k l_c$, $k$ being the wave number of the incident wave, will be referred to as the {\em phase parameter}.

ii) Random heights $V_r$ ($r=1,\cdots n$). 
The $n$ heights $V_r$ are statistically independent of one another; the $n$ distributions are uniform, with zero average, and identical to one another.


The same model has been analyzed later in Ref. \cite{herrera-izrailev-et-al}, using a mapping to a ``classical phase space" and iterating that map.

Systems with similar characteristics have been studied in the past and denoted as periodic-on-average systems, and authors would speak of Kronig-Penney-like models
(see, e.g., Refs. \cite{lifshitz,deych_et_al_1998,mcgurn_et_al_1993,freilikher_et_al_1995,erdos_1982,
gasparian_et_al_1988}).
E. g., they study models where all 1D states are localized, but one group shows regular  Anderson behavior, and a second group, related to gap states, has non-universal properties \cite{deych_et_al_1998}.
Also, the localization length is found to be very small in the gaps and much larger in the bands \cite{mcgurn_et_al_1993}.                    
In Ref. \cite{freilikher_et_al_1995}, the surprising result is found that the transmission coefficient for frequencies associated with the gap in the band structure of the periodic system increases with increasing disorder, for sufficiently weak disorder.

In the problem to be studied in the present paper 
(along the lines of the model outlined above),
we elaborate on previous investigations on disordered systems which are periodic on average, and carry on the following analysis.

i) We also find {\em two regimes} with different {\em localization properties}, 
whose ``evolution" we study in great detail as function of $\delta$ 
(for fixed $l_c$, this means as function of the incident momentum $k$) and $n$. 

ii) We study in detail the {\em transition} between the two regimes; interestingly, in the transition region the problem exhibits interference fringes that give an oscillatory behavior.

iii) We can perform such a detailed study thanks to the fact that we are able to provide an {\em exact theoretical solution} for the average resistance of the system. 
We verify this exact solution by means of {\em computer simulations}.

iv) In addition to the exact solution, we also provide a more {\em qualitative} analysis, based on: \\
a) {\em perturbation theory}, that gives a better physical insight, and   \\
b) the behavior of a finite stretch of a {\em periodic Kronig-Penney model}. 

To carry on this program, the physical quantities we study are the Landauer resistance of the chain \cite{landauer} and its 
Landauer-B\"uttiker conductance \cite{buettiker} $(e^2/h)T$ ($R$ and $T$ being the reflection and transmission coefficients of the chain) averaged over an ensemble of realizations, as functions of the number of scatterers $n$ and the phase parameter 
$\delta$.  

The point of view adopted in the present paper is very much oriented towards condensed matter, although the results are actually much more general, as they have to do with wave propagation.
We may mention that in the domain of ultracold atoms, Anderson localization has been studied and, more impressively, localized matter waves
--in a Bose-Einstein condensate-- have been observed
(see, e. g., \cite{billy_et_al_2008,roati_et_al_2008,chabe'_et_al_2008,lugan_et_al2009}).
The potential considered is a ``speckle potential", an example of a correlated disorder with correlation length $\sigma_R$.
It is remarkable that a transition is observed for $k\sigma_R \sim 1$, reminiscent of the transition for $kl_c \approx \pi$ that we observe in our model: it is as if our ``steps" could be considered as a potential completely correlated for distances smaller than $l_c$ and completely uncorrelated for distances larger than $l_c$.

The paper is organized as follows. 
In the next section we describe the theoretical model using the transfer-matrix technique.
Section \ref{av. landauer} studies the {\em exact} theoretical results for the average Landauer resistance $R/T$ of the chain, as well as the results of computer simulations. 
We first discuss the average Landauer resistance as a function of the number of scatterers $n$ for fixed values of $\delta$, 
a novel feature of these results being their oscillatory behavior. 
We develop a perturbation theory for values of $\delta$ not too close to $\pi$, which gives a qualitative understanding of the oscillations.
We then discuss the average Landauer resistance as a function of $\delta$ for
fixed $n$. 
The remarkable fact is that we observe the ``formation of a gap" very close to 
$\delta=\pi$ (this region will be designated as $\delta \approx \pi$).
In Sec. \ref{<T>} we perform a similar study for the average transmission coefficient of the chain.
In this case, the theoretical results are subject to a number of approximations and are compared with computer simulations, the agreement between both being excellent.
Just as in the case of the resistance, salient features of the results are, on the one hand, their oscillatory behavior and, on the other, the formation of the gap observed for $\delta \approx \pi$.
In Sec. \ref{discussion} we present a more qualitative explanation of the formation of the gap, based on: i) perturbation theory, and ii) the analogy with a finite stretch of a periodic Kronig-Penney model.
We finally conclude in Sec. \ref{concl}.
A number of appendices are added in order not to interrupt the presentation in the main text.

\section{The theoretical model}
\label{theory}

\begin{figure}[t]
\epsfig{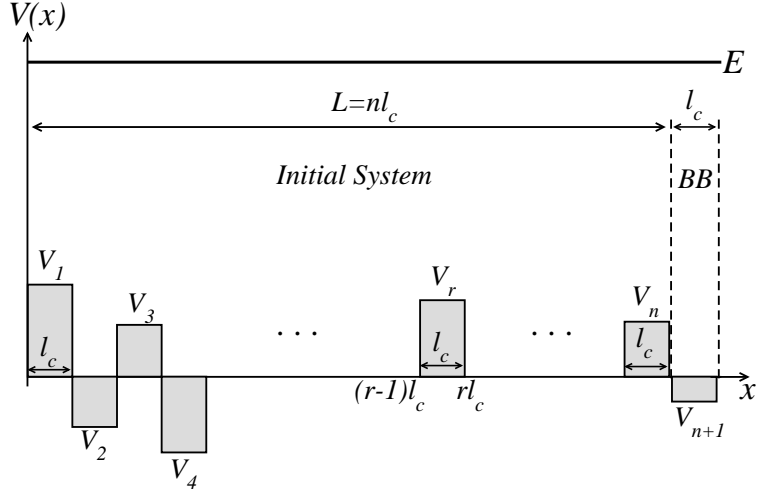}
\caption{
\footnotesize{
Schematic representation of an array of $n$ steps of random height $V_r$ 
($r=1,\cdots n$) possessing a fixed spatial width $l_c$.  
The incident energy $E$ is taken larger than all the $|V_r|$ 's.
Also indicated is the ``initial system" with $n$ scatterers and the addition of the BB consisting of the $(n+1)$-st scatterer.
}
}
\label{rand_steps}
\end{figure}

\begin{figure}[h]
\centering
\epsfig{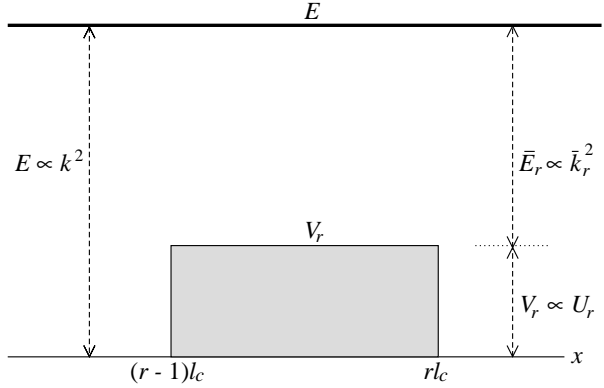}
\caption{
\footnotesize{
Schematic representation of the $r$-th scatterer of the chain, for the case of a barrier.
It has a fixed spatial width $l_c$ and height $V_r$.
}
}
\label{barrier}
\end{figure}

In this section we give a theoretical treatment of the 1D system whose potential, represented schematically in Fig. \ref{rand_steps}, was described in the Introduction.
The $r$-th scatterer of the chain is shown in Fig. \ref{barrier} for the case of a barrier, $V_r>0$; the definitions given below and in the figure also apply to a well, letting $V_r<0$.
In the region of the barrier, the energy $\bar{E}_{r}$ and the wave number $\bar{k}_{r}$ are given by
\begin{subequations}
\begin{eqnarray}
&\hspace{0.37cm}\bar{E}_{r}=E-V_{r},
\label{k_l_j} \\
&\hspace{0.0cm}\left(\bar{k}_{r}\right)^{2}=k^{2}-U_{r},
\label{h_f_d}
\end{eqnarray}
\label{q_w_q_s}
\end{subequations}
where
\begin{equation}
U_{r}=\dfrac{2mV_{r}}{\hbar ^{2}},
\hspace{1cm}
k^{2}=\dfrac{2mE}{\hbar ^{2}}.
\label{U,k2}
\end{equation}
Notice that $k$ is the wavenumber in the absence of barriers.
We also introduce the dimensionless parameter
\begin{equation}
\hspace{0.8cm}y_{r}=U_{r}l_{c}^{2}=\dfrac{U_{r}}{k^{2}}\left(kl_{c}\right)^{2}\equiv \dfrac{U_{r}}{k^{2}}\delta ^{2},
\label{sogamoso}
\end{equation}
as a convenient measure of the intensity of the step potential.

The transfer matrix for the $r$-th scatterer has the structure
\begin{equation}
\textbf{\textit{M}}_{r}=
\begin{bmatrix}
\alpha_{r} & \beta_{r} \\ 
\beta_{r}^{*} & \alpha_{r}^{*} 
\end{bmatrix},
\label{Mr}
\end{equation}
with the condition $|\alpha_{r}|^2-|\beta_{r}|^2 =1$, so that it fulfills the properties of flux conservation and time-reversal invariance \cite{mello-kumar}.
For an incident energy $E$ above a barrier ($ 0<y_{r}<\delta ^{2} $), or for arbitrary $E$ in the case of a well, we find
\begin{subequations}
\begin{eqnarray}
&&\alpha_{r}=e^{-i\delta}\left[\cos\left(\sqrt{\delta^{2}-y_{r}}\right)+i\dfrac{2\delta^{2}-y_{r}}{2\delta\sqrt{\delta^{2}-y_{r}}}\sin\left(\sqrt{\delta^{2}-y_{r}}\right)\right]
\equiv \tilde{\alpha}_{r} \;,
\label{alphar} \\
\nonumber \\
&&\hspace{0cm}
\beta_{r}=-ie^{-i(2r-1)\delta}\dfrac{y_{r}}{2\delta\sqrt{\delta^{2}-y_{r}}}\sin\left(\sqrt{\delta^{2}-y_{r}}\right)
\equiv -ie^{-i(2r-1)\delta}  \tilde{\beta}_{r} \; ,
\label{betar}
\end{eqnarray}
\label{alphar,betar}
\end{subequations}
where the quantities $\tilde{\alpha}_{r}$ and $\tilde{\beta}_{r}$ are independent of the ``running-phase" factor $\exp(-2ir\delta)$.
The transfer matrix associated with a chain containing $n$ (non-overlapping) steps will be denoted by (lower indices refer to individual scatterers)
\begin{subequations}
\begin{eqnarray}
\textbf{\textit{M}}^{(n)}
&=& \textbf{\textit{M}}_{n}
\cdots\textbf{\textit{M}}_{r}\cdots\textbf{\textit{M}}_{2}
\textbf{\textit{M}}_{1}  \\
&=& 
\begin{bmatrix}
\alpha^{(n)} & \beta^{(n)} \\
\left(\beta^{(n)}\right)^{*} & \left(\alpha^{(n)}\right)^{*} 
\end{bmatrix} 
= 
\left[
\begin{array}{cc}
e^{i \varphi^{(n)}} & 0 \\
0 & e^{-i \varphi^{(n)}}
\end{array}
\right]
\left[
\begin{array}{cc}
\sqrt{1+ \lambda^{(n)}} & \sqrt{\lambda^{(n)}} \\
\sqrt{\lambda^{(n)}} & \sqrt{1+ \lambda^{(n)}}
\end{array}
\right]
\left[
\begin{array}{cc}
e^{i \psi^{(n)}} & 0 \\
0 & e^{-i \psi^{(n)}}
\end{array}
\right]
\; .
\nonumber \\
\end{eqnarray}
\label{M(n)}
\end{subequations}
Here, $\varphi^{(n)}$ and $\psi^{(n)}$ are phases, and $\lambda^{(n)}$ is the ``radial" parameter in the polar representation of the transfer matrices \cite{mello-kumar}.

Quantities of particular physical interest are the Landauer resistance
\cite{landauer} $\lambda^{(n)}$ of the chain 
\begin{subequations}
\begin{eqnarray}
\lambda^{(n)} 
&=& |\beta^{(n)}|^2 
= \frac{R^{(n)}}{T^{(n)}}
\label{lambda(n)} 
\end{eqnarray}
and its dimensionless Landauer-B\"uttiker conductance given by the transmission coefficient
\begin{eqnarray}
T^{(n)} &=& \frac{1}{1+\lambda^{(n)}}\; .
\label{T(n)}
\end{eqnarray}
\label{lambda(n),T(n)}
\end{subequations}

The ensemble of chains described in the Introduction is defined by assuming that the $y_{r}$'s ($r=1,\cdots,n$) are statistically independent of one another, each being uniformly distributed in the interval 
$\left(-y_{0},y_{0}\right)$.
This is equivalent to saying that, for fixed $\l_c$, each $U_r$ is uniformly distributed in the interval $\left(-U_{0},U_{0}\right)$, with
$y_{0}\equiv U_{0}l_{c}^{2}$.
If each chain is represented as in Eq. (\ref{M(n)}), the ensemble of chains is described by an {\em ensemble of transfer matrices}.

It is relevant here to comment on the dependence of the physical quantities of interest on the parameters that we have introduced.
Notice that, although the transfer matrix for a single scatterer depends, in principle, on the three parameters $E, U_r, l_c$, Eqs. (\ref{alphar,betar}) show that these parameters occur in the combinations $\delta$ and $y_r$.
Thus, for the full chain of $n$ scatterers and a specific realization of disorder, a quantity like the transmission coefficient $T^{(n)}$ depends on the various parameters as
\begin{equation}
T^{(n)} 
=f(\delta, n, y_1, \cdots, y_n) \; .
\label{T(n) parameters}
\end{equation}
Its ensemble average is thus given by
\begin{subequations}
\begin{eqnarray}
\langle T^{(n)}  \rangle
&=& \int \cdots \int f(\delta, n, y_1, \cdots, y_n)
\; p_{y_0}(y_1) \cdots p_{y_0}(y_n) \; dy_1 \cdots dy_n 
\label{<T> a}  \\
&=& F(\delta, n, y_0) \; ,
\label{<T> b} 
\end{eqnarray}
\label{<T>}
\end{subequations}
which is seen to depend on the three parameters $\delta, n$ and $y_0$ only.

\section{Average Landauer resistance}
\label{av. landauer}

We assume that the original system of $n$ scatterers is extended with the addition of one scatterer, to be called a {\em ``building block"} (BB), as shown in 
Fig. \ref{rand_steps}.
The resulting transfer matrix is given by 
\begin{equation}
\textbf{\textit{M}}^{(n+1)} 
=\textbf{\textit{M}}_{n+1} \textbf{\textit{M}}^{(n)}.
\label{M(n+m)=M(m)M(n)}
\end{equation}
From this combination rule we find the recursion relation for Landauer's resistance of the chain, averaged over the ensemble, given in 
App. \ref{recursion}, Eqs. (\ref{rec.rel. 1,2}).
Notice that Eqs. (\ref{rec.rel. 1,2}) couple the average resistance of the chain, $\langle\vert\beta^{(n)}\vert^{2}\rangle$, to the quantity
$\langle\alpha^{(n)}\beta^{(n)}\rangle$.
The recursion relations (\ref{rec.rel. 1,2}) are exact, and thus take into account all multiple scattering processes occurring in the chain.

Eqs. (\ref{rec.rel. 1,2}) can be written as a recursion relation for the quantities
\begin{subequations}
\begin{eqnarray}
A(n)
&=& 1+2\langle\vert\beta^{(n)}\vert^{2}\rangle,
\label{A(n)} \\
b(n)
&=& e^{2in\delta}\langle \alpha^{(n)}  \beta^{(n)}\rangle,
\label{b(n)}
\end{eqnarray}
\label{A(n),b(n)}
\end{subequations}
which is given explicitly in Eq. (\ref{z(n+1)=Mz(n) 1}).
Using the definition
\begin{equation}
z(n) =  \left[\frac{A(n)}{2}, \frac{i b(n)}{\sqrt{2}}, 
-\frac{i b^{*}(n)}{\sqrt{2}}\right] ^T \; ,
\label{z}
\end{equation}
($T$ meaning transpose), we see that Eq. (\ref{z(n+1)=Mz(n) 1}), in turn, has the simple structure
\begin{equation}
z(n+1) = \Omega_{y_0}(\delta) z(n) \; .
\label{z(n+1)=Mz(n) 2}
\end{equation}

We have assumed that all the individual scatterers are equally distributed, so that the various BB averages can be evaluated for the first scatterer.
In Eq. (\ref{z(n+1)=Mz(n) 2}), $\Omega_{y_0}(\delta)$ 
is the $3 \times 3$ matrix appearing on the right-hand side of Eq. (\ref{z(n+1)=Mz(n) 1}).
The matrix $\Omega_{y_0}(\delta)$, which depends on $y_0$ and $\delta$, will be denoted by $\Omega$, for short,  when no confusion arises.
The various BB averages appearing in $\Omega$ are to be evaluated using the expressions of Eqs. (\ref{alphar,betar}).

The matrix $\Omega$ we have defined is {\em complex symmetric} and 
{\em independent of $n$}. Thanks to this last property, the solution of 
Eq. (\ref{z(n+1)=Mz(n) 2}) for arbitrary $n$ can be written as
\begin{subequations}
\begin{eqnarray}
z(n) &=& \Omega ^n z(0) 
\label{sol. z(n) 1} \\
z(0)&=&[1/2,0,0]^T \;.
\label{z(0)}
\end{eqnarray}
\label{z(n),z(0)}
\end{subequations}
This is done in detail in App. \ref{diagonalizing Omega}, through the diagonalization of the matrix $\Omega$.

For the average Landauer resistance [see Eqs. (\ref{lambda(n)}) and (\ref{A(n)})] we obtain
\begin{equation}
\langle \lambda^{(n)} \rangle
= \frac12[A(n)-1].
\label{lambda(n) 1}
\end{equation}

\subsection{Average Landauer resistance in regime I, as function of the number of scatterers $n$}
\label{<R/T> regime I vs n}

Assume $\delta$ is far from $\pi$. E.g., for $\delta=\pi/2$, the three unperturbed eigenvalues of $\Omega_0$ are 
$\{\mu_1^{(0)},\mu_2^{(0)},\mu_3^{(0)} \} = \{1, -1, -1\}$.
We call {\em regime I} the region in which $\{ \mu_2^{(0)},\mu_3^{(0)} \}$ are far away from $\mu_1^{(0)}$, so that they may be considered effectively decoupled when we turn on a weak interaction, $y_0^2 \ll 1$.
We then restrict ourselves to the $1\times 1$ block of $\Omega$ in 
Eq. (\ref{z(n+1)=Mz(n) 1}) consisting of the 11 matrix element, and write the solution, Eq. (\ref{sol. z(n) 1}), as
\begin{equation}
A(n) \approx \Omega_{11}^n A(0)
=\left(1+2
\langle\vert\beta_{1}\vert^{2}\rangle\right)^{n}=e^{2n\frac{1}{2}\ln \left(1+2
\langle\vert\beta_{1}\vert^{2}\rangle\right)}\equiv e^{2nl_{c}/\ell},
\label{A(n) 2}
\end{equation}
which defines the parameter $\ell$, to be interpreted below.
Eq. (\ref{A(n) 2}) is the well known exponential behavior found by Landauer \cite{landauer}, where, in the present case,
\begin{equation}
\dfrac{l_{c}}{\ell}=\frac{1}{2}\ln \left(1+2
\langle\vert\beta_{1}\vert^{2}\rangle\right),
\label{mfp 1}
\end{equation}
where $\beta_1$ refers to the first scatterer.
In the WSL, $\langle |\beta _{1} |^{2}\rangle =\langle R_{1}/T_{1}\rangle \ll 1$, and we can write
\begin{subequations}
\begin{eqnarray}
\dfrac{l_{c}}{\ell}
&\approx&
\langle\vert\beta_{1}\vert^{2}\rangle =\langle R_{1}/T_{1}\rangle \approx\langle R_{1}\rangle \; ,
\label{lc/l}  
\end{eqnarray}
so that
\begin{eqnarray}
\dfrac{1}{\ell} &\approx& \dfrac{\langle R_{1}\rangle}{l_{c}} \; .
\label{mfp 2}
\end{eqnarray}
\end{subequations}
Thus $1/\ell$ is, approximately, the reflection coefficient per unit length, that we shall identify with the inverse of the {\em mean free path} (mfp)
\cite{froufe_et_al_2007}, which, in the present 1D problem, is of the order of the
localization length.

Explicitly, Landauer's resistence for the chain consisting of $n$ scatterers in regime I takes the form
\begin{equation}
\langle\vert\beta^{(n)}\vert^{2}\rangle =\dfrac{1}{2}\left(e^{2nl_{c}/\ell} -1\right).
\label{landauer(n)}
\end{equation}
Using Eq. (\ref{betar}), we can express 
$\langle |\beta _{1} |^{2}\rangle$ appearing in (\ref{mfp 1}) as function of
$\delta $ and $ y_{0}$ as
\begin{equation}
\langle |\beta _{1} |^{2}\rangle =\left\langle \dfrac{y_{1}^{2}}{4\delta ^{2}\left(\delta ^{2}-y_{1} \right)}\sin ^{2}\left(\sqrt{\delta ^{2}-y_{1}} \right) \right\rangle .
\label{<beta1^2> 1}
\end{equation}
Although this average can be computed analytically and expressed in terms of cosine-integral functions, in future calculations it will be more convenient to compute it numerically.
However, it is worth noticing that in the WSL it can be expanded in powers of $y_{0}/\delta ^{2}$, giving the rather compact and transparent expression
\begin{equation}
\langle |\beta _{1} |^{2}\rangle = \dfrac{l_{c}}{\tilde{\ell}} + O\left(\dfrac{y_{0}}{\delta ^{2}}\right)^{4}, \qquad\qquad \dfrac{l_{c}}{\tilde{\ell}}=\dfrac{y^{2}_{0}}{12\delta^{4}}\sin^{2}\delta .
\label{<beta1^2> 2}
\end{equation}
\begin{figure}[h]
\centering
\epsfig{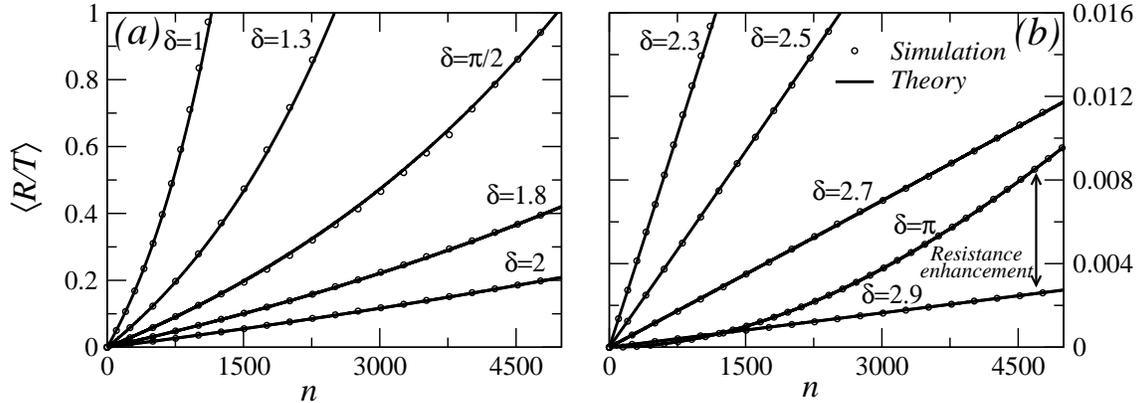}
\caption{
\footnotesize{
Theory and computer simulations for $\langle\left(R/T\right)^{(n)}\rangle$ 
as a function of $n$ for the 1D system described in the text.
For the numerical simulation, an ensemble of $10^{4}$ realizations was used. 
Results are shown in regime I for a number of $\delta$'s:
(a) $1< \delta <2$, (b)  $ 2.3 < \delta < 2.9$;
$\delta=\pi$, from regime II, is also shown.
We can observe the localization properties described in the text.
We chose the parameter $y_{0} =$ 0.09. 
The error bar due to the finite sample size is very small and is not indicated in the figure: e.g., for 
$\delta =\pi/2$ and $n=5000$, the error is $\sim 10^{-2}$.
}
}
\label{land_res_reg_Ia}
\end{figure}
Notice that in the present problem {\em the mfp depends on the phase parameter} $\delta$.

We now compare the theoretical result of Eq. (\ref{landauer(n)}) with numerical simulations.
In the WSL we have $y_{0}/\delta ^{2} \ll 1$;
we fix $y_0 = 0.09$ and consider $\delta$ in the interval $(1,2.9)$.
Figure \ref{land_res_reg_Ia} shows the theoretical results and numerical simulations for the average Landauer resistance as functions of the length $n$ of the chain, for various values of $\delta$ in the above interval: 
the agreement is excellent, indicating that the decoupling leading to the simple equation (\ref{landauer(n)}) for the resistance, as well as the expression (\ref{<beta1^2> 2}) for the mfp are very good approximations.

The results indicate the {\em tendency of the system to delocalize, with a corresponding increase in the mfp, as the phase parameter $\delta$ increases towards} $\pi$.

\subsection{Average Landauer resistance in regime II, as function of the number of scatterers $n$}
\label{<R/T> regime II vs n}

In the region $2.9 \lesssim \delta \lesssim 3.4$, 
$\{ \mu_2^{(0)},\mu_3^{(0)} \}$ are not far enough away from $\mu_1^{(0)}$ to be effectively decoupled. 
We shall see that {\em a novel behavior shows up as a consequence of the coupling}. 

\subsubsection{The behavior of the average resistance for $\delta = \pi$}
\label{delta=pi}

For $\delta = \pi$, the three $\mu_a^{(0)}$ are degenerate and equal 
to 1.
In this case, and for weak scattering, i.e., $y_0 \ll 1$, $\Omega$ takes the approximate form given in Eqs. (\ref{Omega,Omega-red y0<<1}).

Theoretical results (obtained diagonalizing $\Omega$ of 
(\ref{Omega,Omega-red y0<<1}) numerically) and computer simulations for the average Landauer resistance for $\delta=\pi$ are also shown in 
Fig. \ref{land_res_reg_Ia} as a function of $n$.
The excellent agreement between the two results indicates that writing $\Omega$ as in Eqs. (\ref{Omega,Omega-red y0<<1}) is a good approximation.
What we learn is that the system is {\em less delocalized for $\delta = \pi$ than for neighboring values of} $\delta$:
i.e., {\em the tendency to delocalize as $\delta$ moves towards $\pi$ is reversed for} $\delta = \pi$,
where we notice an enhancement of the average resistance.

\subsubsection{Perturbation theory for $\delta$ not too close to $\pi$}
\label{pert. theory}

For $\delta$ not too close to $\pi$, so that the unperturbed eigenvalues do not become degenerate, we may use perturbation theory (PT) in the parameter $y_0$ to find approximate expressions for the eigenvalues and eigenvectors of the matrix $\Omega$ appearing in the recursion relation (\ref{z(n+1)=Mz(n) 2}), as is 
briefly discussed in App. \ref{pert. theo.}.
We write 
$\Omega_{y_0}(\delta) = \Omega_0 (\delta) + \Delta \Omega_{y_0}(\delta)$ as in Eqs. (\ref{M=M0+DeltaM}), and consider 
$\Delta \Omega_{y_0}(\delta)$ as a perturbation;
the latter contains the BB expectation values appearing in 
Eq. (\ref{z(n+1)=Mz(n) 1}).
The perturbation can be calculated analytically in leading order in $y_0^2$, as we did with $\langle |\beta _{1} |^{2}\rangle$, Eq. (\ref{<beta1^2> 2}).
However, just as we mentioned right below Eq. (\ref{<beta1^2> 1}),
it is convenient to have an exact expression for these BB quantities, so as to have a better control on the perturbation expansion:
they were thus evaluated numerically.
\begin{figure}[h]
\centering
\epsfig{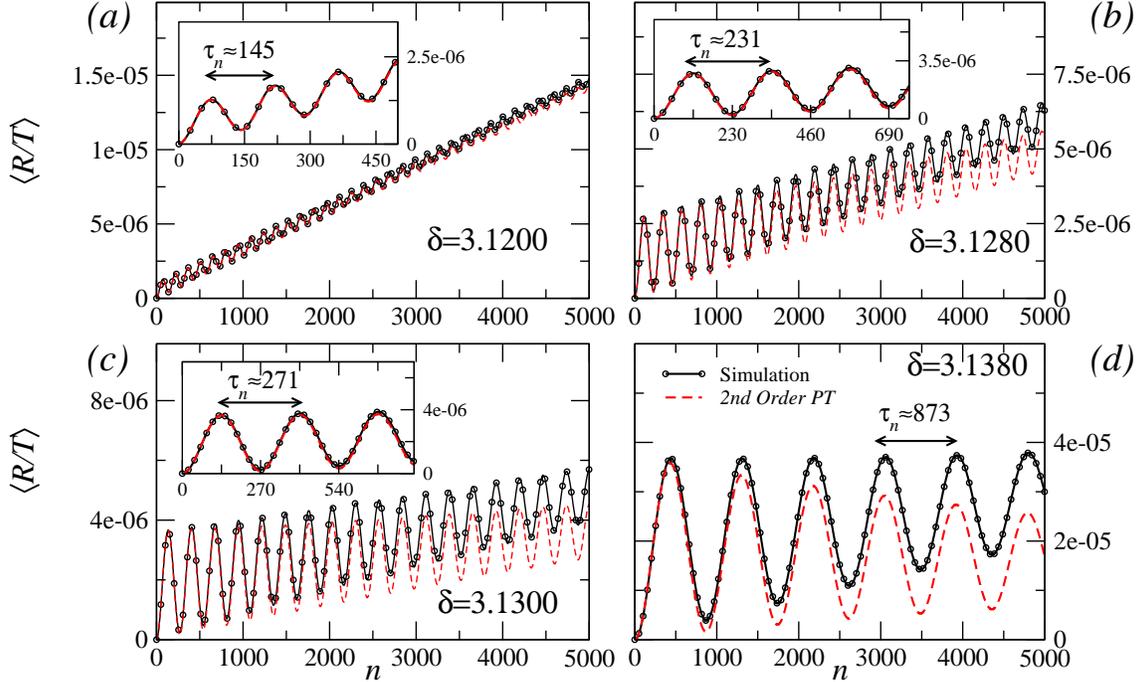}
\caption{
\footnotesize{
Perturbation theory and computer simulations for $\langle (R/T)^{(n)} \rangle$ as a function of $n$, for $\delta=3.1200, 3.1280, 3.1300, 3.1380$ and $y_0=0.09$.
Perturbation theory was carried out up to second order in $\Delta \Omega$ in the eigenvalues and the eigenvectors.
The description is reasonable, especially in the first three cases 
(a,b and c); in the fourth case (d) the agreement deteriorates, as $\delta$ is too close to $\pi$.
Notice the {\em oscillatory behavior} as a function of $n$.
The insets in (a), (b) and (c) are a zoom of the results for the first few oscillations.
The estimate of the period from Eq. (\ref{tau-n}) is indicated in each panel and
is consistent with the numerical data.
}
}
\label{pert_theo_land_res_d_3.1380_y0_0.09_0.01}
\end{figure}

Fig. \ref{pert_theo_land_res_d_3.1380_y0_0.09_0.01} shows the results of perturbation theory and simulations for the average Landauer resistance as a function of $n$, for four values of $\delta$.
A salient novel feature of these results is their 
{\em oscillatory behavior as a function of} $n$;
in the case of scatterers with a vanishing size and for a fixed wavelength as in previous studies \cite{froufe_et_al_2007}, oscillations with the present origin were absent.
This behavior can be understood as follows.
From Eq. (\ref{A(n) 3}), $A(n)$ has the structure
\begin{equation}
A(n) 
\sim 
A_1 e^{n \ln (1+ \Delta \mu_1)}  
+\left[ A_2 e^{n \ln (e^{2i\delta}+\Delta \mu_2)} + cc \right] \; ,
\end{equation}
where $A_1$, $A_2$ are constants independent of $n$ and
 $\Delta \mu_i = \mu_i - \mu_i^{(0)}$.
For $\delta = \pi + \epsilon$ and neglecting $\Delta \mu_i$,
\begin{equation}
e^{n\ln e^{2i\delta}} = e^{n2i\delta}
= e^{2in(\pi + \epsilon)}
=e^{2in\epsilon} \; .
\label{oscillations}
\end{equation}
This result oscillates with $n$, with a period $\tau_n$ that satisfies 
$2 \epsilon \tau_n  = 2 \pi$, so that, for $\delta fixed$, we estimate
\begin{equation}
\tau_n \sim \frac{\pi}{\epsilon} \; .
\label{tau-n}
\end{equation}
This estimate for the period $\tau_n$ is independent of $y_0$, it decreases as $\delta$ moves away from $\delta = \pi$, and is consistent with the results of Fig. \ref{pert_theo_land_res_d_3.1380_y0_0.09_0.01}.

\subsubsection{Exact solution for $\delta$ very close to $\pi$ 
($\delta \approx \pi$)}
\label{delta very close to pi}

\begin{figure}[h]
\centering
\epsfig{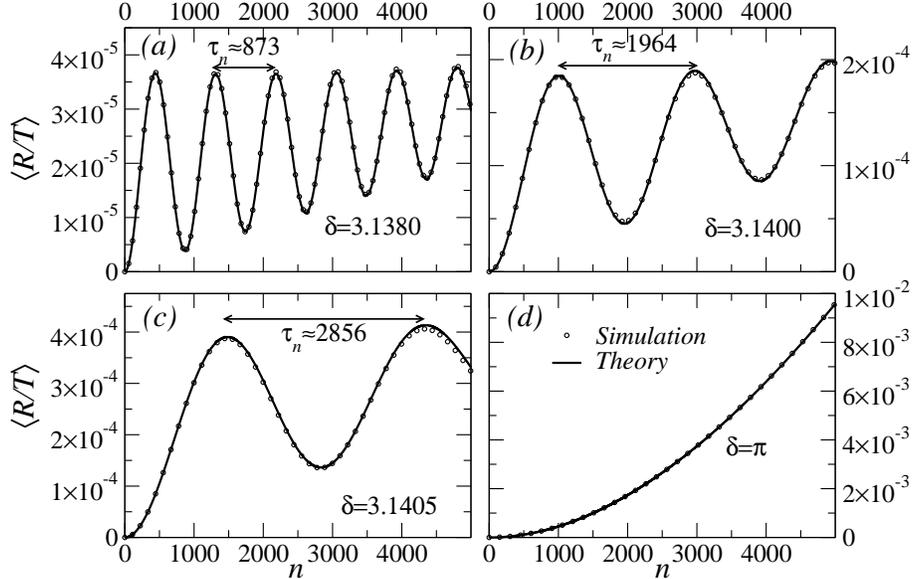}
\caption{
{\small
Numerical simulations and analytical solution for
$\langle (R/T)^{(n)} \rangle$ vs. $n$, obtained diagonalizing numerically the matrix $\Omega_{y_0}(\delta)$, for 
$y_0=0.09$ and for four values of 
$\delta \approx \pi$:
(a)-(c) $\delta =3.1380, 3.1400, 3.1405$;
(d) $\delta = \pi$.
The analytical solution is essentially exact.
The results are symmetric around $\delta = \pi$ in the vicinity of this value.
As a verification of the theoretical results, also shown are computer simulations using an ensemble of $10^4$ realizations.
For $\delta = \pi$, the statistical error bar is smaller than $10^{-5}$ and is not indicated.
The estimate from Eq. (\ref{tau-n}) of the period $\tau_n$ of the oscillations is indicated in each panel and is consistent with the numerical simulations and the analytical results.
}
}
\label{land_res_d_pi-3.1380_y0_0.01}
\end{figure}

If $\delta$ is very close to $\pi$, perturbation theory fails and $\Omega$ has to be diagonalized exactly.
This has been done for a number of cases, shown in 
Fig. \ref{land_res_d_pi-3.1380_y0_0.01}.
The analytical results are a plot of the solution for the average Landauer resistance given in Eqs. (\ref{lambda(n) 1}), in which the matrix $\Omega_{y_0}(\delta)$ was diagonalized numerically.
These results, which are essentially exact,
have been verified with the aid of computer simulations, also shown in the figure.
Notice again the oscillatory behavior of the resistance as a function of $n$: the period $\tau_n$ of the oscillations decreases as $\delta$ goes away from $\pi$, as we already noted in relation with Eq. (\ref{tau-n}).


\subsection{Average Landauer resistance in regimes I and II, as function of $\delta$ for fixed $n$}
\label{landauer reg I,II vs delta}

\begin{figure}[h]
\centering
\epsfig{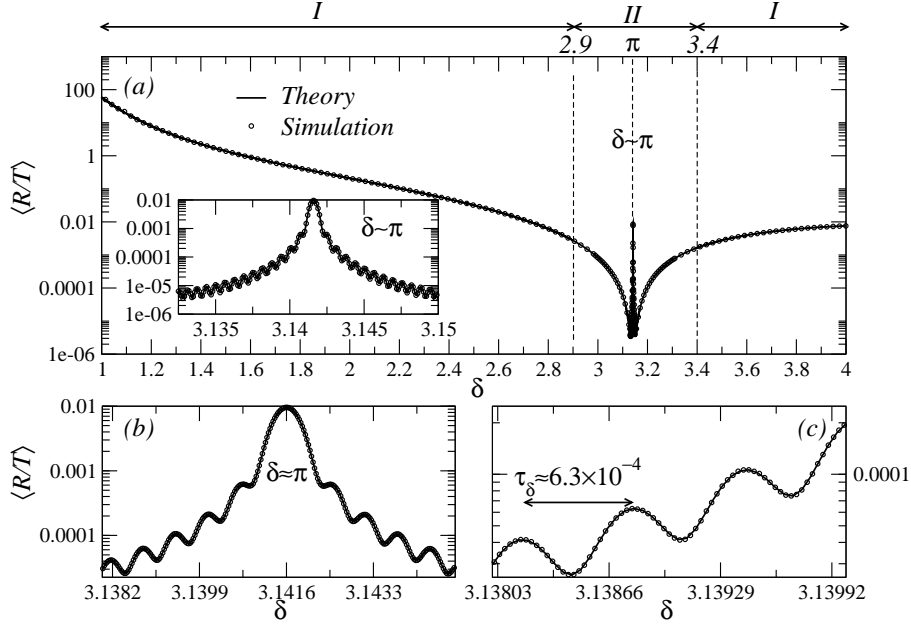}
\caption{
{\footnotesize
Theory and numerical simulations for $\langle (R/T)^{(n)} \rangle$ as a function of $\delta$, for a chain of $n=5000$ scatterers and for $y_0 = 0.09$, in regimes I and II.
The simulations use $10^5$ realizations.
In regime II, Eqs. (\ref{lambda(n) 1}) and  (\ref{A(n) 1})
were employed, diagonalizing numerically the matrix $\Omega$; the computer simulations 
(with a statistical error bar $\sim 10^{-5}$ for $\delta=\pi$) constitute a verification of the theoretical results.
Well inside regime II, we observe the dramatic enhancement of the average resistance by nearly {\em three orders of magnitude}, explained in the text.
The inset in panel a), and panels b) and c) show this latter region in greater detail.
Notice the {\em oscillatory behavior of the average Landauer resistance as a function of $\delta$ for fixed $n$}.
The period of the oscillations can be estimated from Eq. (\ref{tau-delta}) as $\tau_{\delta} \sim 6 \cdot 10^{-4}$, which is consistent with what we observe in the figure
(in spite of $\delta$ being quite close to $\pi$).
}
}
\label{land_res_reg_Ia_6}
\end{figure}
We gain a global picture of the two regimes if we study the behavior of the average resistance $\langle R/T \rangle$ for a fixed length $n$ of the chain, as a function of the phase parameter $\delta$.

Fig. \ref{land_res_reg_Ia_6} shows the analytical results for $n=5000$ scatterers and $1< \delta <4$, covering regimes I and II.
We observe in Fig. \ref{land_res_reg_Ia_6}a that the average resistance decreases as $\delta$ moves towards $\pi$, in agreement with the picture we have described of the system becoming more delocalized.
The theoretical curve corresponding to regime I 
($1< \delta <2.9$ and $\delta >3.4$) was again obtained from Eq. (\ref{landauer(n)}), the comparison with the simulation being excellent.

In Regime II, the matrix $\Omega$ was diagonalized as before. 
These results were verified by computer simulations, also shown in 
Fig. \ref{land_res_reg_Ia_6}.
In agreement with the earlier discussion of Fig. \ref{land_res_reg_Ia},
we observe that well inside regime II the propensity of the average resistance to decrease as $\delta$ moves towards $\pi$ is reversed,
indicating the {\em formation of a gap}.
A discussion of the physical interpretation of this phenomenon will be given in 
Sec. \ref{discussion}.

The inset in panel (a) of Fig. \ref{land_res_reg_Ia_6} exhibits an {\em oscillatory behavior of the average Landauer resistance as a function of $\delta$ for fixed $n$}.
Again, this effect was not there in earlier studies in which the scatterers had a vanishing size.
We can estimate the period from the perturbative result given in 
Eq. (\ref{oscillations}) as 
\begin{equation}
\tau_{\delta} \sim \frac{\pi}{n} \; ,
\label{tau-delta}
\end{equation}
if $\delta$ is not too close to $\pi$.

\section{Average transmission coefficient (Landauer-B\"uttiker conductance)}
\label{<T>}

\subsection{Average transmission coefficient in regime I as function of the number of scatterers $n$}
\label{<T> regime_I vs n}

In this section we analyze the average transmission coefficient 
$\langle T \rangle$ in regime I for the chains that we have been studying.
Since for this quantity we have not succeeded in finding a recursion relation of the type obtained in Eq. (\ref{z(n+1)=Mz(n) 2}) for the average Landauer resistance, we resort to an approximate treatment.

From Eq. (\ref{A(n) 2}), valid in regime I, and treating $n$ approximately as a continuous variable, we write
\begin{equation}
\frac{\partial A(n)}{\partial n} 
\approx 2 \frac{l_c}{\ell} A(n).
\label{diff eqn A(n)}
\end{equation}
In terms of the polar representation \cite{mello-kumar} already employed in previous sections, i.e.,
$\lambda_r = |\beta_r|^2$ for the $r$-th scatterer and 
$\lambda^{(n)}=|\beta^{(n)}|^2$ for the chain consisting of $n$ scatterers,
Eq. (\ref{diff eqn A(n)}) becomes
\begin{equation}
\frac{\partial \langle \lambda \rangle_{s}}{\partial s}
= 1+2 \langle \lambda \rangle_{s} \; ,
\label{evol_eqn_for <lambda>_s}
\end{equation}
where
\begin{equation}
s = nl_c/\ell = L/\ell \; .
\label{s}
\end{equation}
This ``evolution" with $s$ of $\langle \lambda \rangle_{s}$ 
coincides with that found from the evolution equation for the $\lambda$ probability density, $w_{s}(\lambda)$, known as Melnikov's equation
\cite{mello-kumar,melnikov}
\begin{equation}
\frac{\partial w_{s}(\lambda)}{\partial s}
= \frac{\partial}{\partial \lambda} \left[\lambda(1+\lambda)
\frac{\partial w_{s}(\lambda)}{\partial \lambda}\right].
\label{melnikov}
\end{equation}
We {\em propose} the validity of Melnikov's equation for regime I and verify the consequences numerically.
In particular, from this assumption we can find the statistical properties of the transmission coefficient $T$ which, in terms of $\lambda$, can be written as
\begin{equation}
T=\frac{1}{1+\lambda} \; ;
\label{T(lambda)}
\end{equation}
indeed, from Melnikov's equation (\ref{melnikov}), the expression for the $p$-th moment of $T$ can be reduced to quadratures, with the result
\cite{marlos_et_al}
\begin{equation}
\langle T^p \rangle
= \frac{2 {\rm e}^{-\tilde{s}/4}}{\Gamma(p)}
\int_0^{\infty}{\rm e}^{-\tilde{s}t^2}
\left|\Gamma\left(p-\frac12 +it\right)\right|^2  t \; \tanh (\pi t) {\rm d} t  \; ,
\label{<Tp>_Melnikov}
\end{equation}
from which we find the first moment as
\begin{equation}
\langle T \rangle
= 2 {\rm e}^{-\tilde{s}/4}
\int_0^{\infty}{\rm e}^{-\tilde{s}t^2}
\pi t [\tanh (\pi t)/\cosh(\pi t)] {\rm d}t \; .
\label{<T>_Melnikov}
\end{equation}

\begin{figure}[h]
\centering
\epsfig{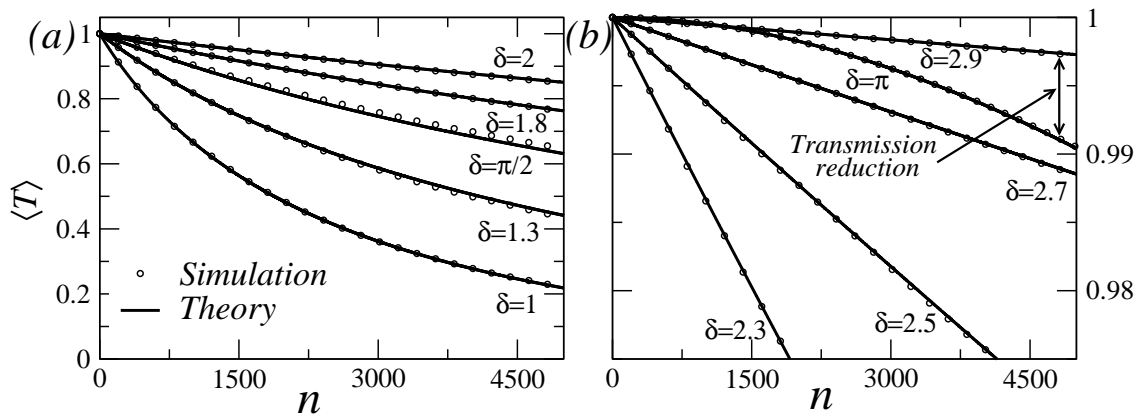}
\caption{
{\footnotesize
Theory and numerical simulations for the average transmission coefficient
$\langle T \rangle$ as a function of the number $n$ of scatterers and for various values of the phase parameter $\delta$ in regime I, in the range: 
(a) $1< \delta < 2$, and (b) $2.3 < \delta < 2.9$,
as in Fig. \ref{land_res_reg_Ia}.
For the simulation, an ensemble of $10^{4}$ realizations was used.
As usual, we chose the parameter $y_{0}=$ 0.09.
The theoretical results, obtained from Eq. (\ref{<T>_Melnikov}), lie on top of the numerical ones. 
The error bar due to the finite sample is not indicated in the figure:
e. g., for $\delta =\pi/2$ and  $n=5000$, the error is $\sim  10^{-2}$.
}}
\label{<T>_regime_I}
\end{figure}

In Fig. \ref{<T>_regime_I} we compare result (\ref{<T>_Melnikov}) with numerical simulations obtained for various values of $\delta$ in regime I as a function of the length $n$ of the chain: the agreement is excellent, indicating that the approximation involved in using Melnikov's equation is reasonable.
The localization properties are consistent with what we observed for the resistance in Fig. \ref{land_res_reg_Ia}:
the transmission {\em reduction} shown in panel (b) is also consistent with the resistance
{\em enhancement} shown in panel (b) of Fig. \ref{land_res_reg_Ia}.

\subsection{Average transmission coefficient in regime II as function of the number of scatterers $n$}
\label{<T> regime_II vs n}

\begin{figure}[h]
\epsfig{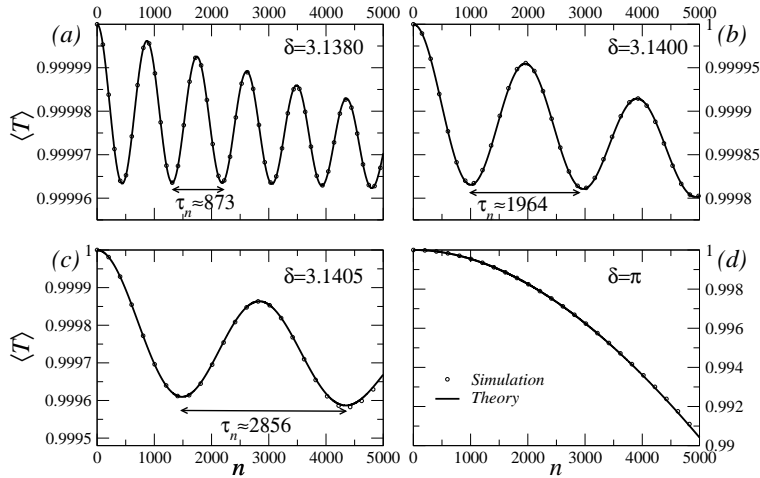}
\caption{
{\footnotesize
The theoretical average transmission coefficient 
$\langle T \rangle$ vs. $n$, obtained from the approximation of 
Eq. (\ref{T(lambda) 1}), for
$y_0=0.09$ and for four values of 
$\delta \approx \pi$:
(a)-(c) $\delta =3.1380, 3.1400, 3.1405$;
(d) $\delta = \pi$
(as in Fig. \ref{land_res_d_pi-3.1380_y0_0.01}),
compared with numerical simulations.
The agreement is excellent, suggesting that the approximation 
of Eq. (\ref{T(lambda) 1}) is justified.
The estimate from Eq. (\ref{tau-n}) of the period $\tau_n$ of the oscillations is indicated in each panel and is consistent with the analytical results and the numerical simulations.
}}
\label{<T>_vs_n for fixed deltas reg. II}
\end{figure}

In regime II, the theoretical analysis uses the approximation 
(see Eq. (\ref{T(n)}))
\begin{equation}
\langle T \rangle
\approx 1 - \langle \lambda \rangle \; ,
\label{T(lambda) 1}
\end{equation}
since $\langle \lambda \rangle \ll 1$
(see Fig. \ref{land_res_reg_Ia_6}), and $\langle \lambda \rangle$
is obtained from the results of the previous section which make use of
the exact recursion relation (\ref{z(n+1)=Mz(n) 2})
and diagonalization of the matrix $\Omega$.
The results, together with numerical simulations, are shown in 
Fig. \ref{<T>_vs_n for fixed deltas reg. II}
for $\delta=\pi$ and very close to $\pi$.
From the excellent agreement we see that our basic approximation, 
Eq. (\ref{T(lambda) 1}), appears justified.

Again, the oscillations shown in Fig. \ref{<T>_vs_n for fixed deltas reg. II} are a novel feature of these results, arising from finite-size scatterers.
The period $\tau_n$ of the oscillations can be taken over from the footnote to 
Fig. \ref{land_res_d_pi-3.1380_y0_0.01} and is consistent with what we observe in Fig. \ref{<T>_vs_n for fixed deltas reg. II}.

\subsection{Average transmission coefficient in regimes I and II, as function of $\delta$ for fixed $n$}
\label{<T> reg. I,II vs delta}

\begin{figure}[h]
\centering
\epsfig{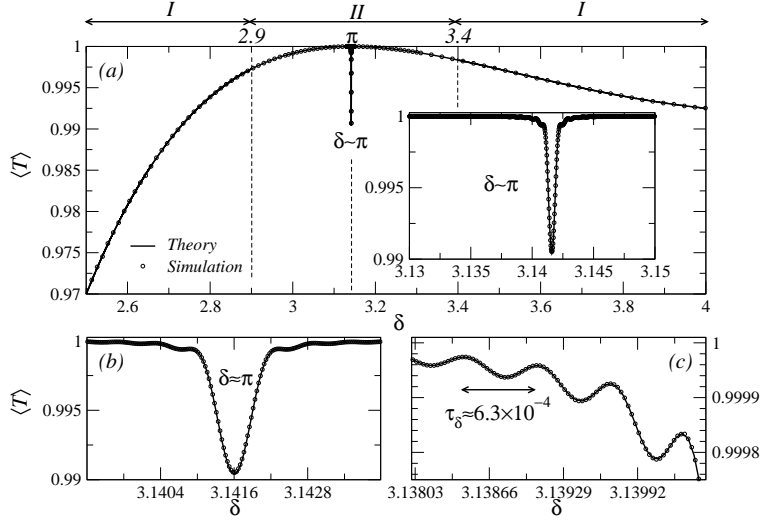}
\caption{
{\footnotesize 
Theoretical results (as described in the text) and numerical simulations for $\langle T \rangle$ vs $\delta$ in Regimes I and II, for a chain of $n=5000$ scatterers and $10^5$ realizations.  
The main figure shows the ``gross-structure" behavior and the dip for 
$\delta \approx \pi$ exhibiting the formation of a band gap, or forbidden region: 
a zoom of the latter is shown in the insets.
The agreement between simulation and theory is excellent.  
Notice the interference fringes in the inset; the period $\tau_{\delta}$ of the oscillations was estimated using Eq. (\ref{tau-delta}) and agrees well with the data (in spite of $\delta$ being quite close to $\pi$).
The statistical error bar for $\delta=\pi$ is $\sim 10^{-5}$.
}
}
\label{<T>_vs_delta}
\end{figure}

Just as we did in the case of the resistance in 
Sec. \ref{landauer reg I,II vs delta}, we now analyze the behavior of the average conductance 
$\langle T \rangle$ for a fixed length $n$ of the chain, as a function of the phase parameter $\delta$.
Fig. \ref{<T>_vs_delta} shows the analytical and numerical results for $n=5000$ scatterers and $2.5< \delta <4$, covering regimes I and II.
In regime I, the analytical results are obtained from 
Eq. (\ref{<T>_Melnikov}), which gives an excellent description of the data.
In regime II, the analytical results are obtained from 
Eq. (\ref{T(lambda) 1}) and $\langle \lambda \rangle$ is extracted from the results of Sec. \ref{av. landauer}.

Fig. \ref{<T>_vs_delta} shows that the average conductance exhibits a 
``gross-structure"  in the form of a ``bump".
For the case of weak scatterers, the system is almost transparent in regime II, and regime I is more localized.
This gross-structure behavior is not entirely surprising. 
A single barrier with fixed width and strength
becomes completely transparent ($T=1$) at the resonance values  
${\bar k}l_c= n\pi$, $n=1,2,\cdots$, where $\bar{k}$ is the wave number in the region of the barrier ($\delta \gtrsim \pi$ for low barriers).  
For a well, $T=1$ at $\delta \lesssim \pi$.  
For a fixed step width and random strength with zero average, and still for $n=1$,
$\langle T \rangle$ reaches a maximum value smaller than unity at $\delta=\pi$. 
As the number of scatterers $n$ increases, the gross structure seen in  $\langle T \rangle$ as a function of $\delta$ is still similar to the above description for one random scatterer, in that
{\em regime II} ($\delta \sim \pi$) shows  
{\em the system to be almost transparent} and less localized than in regime I.

The behavior of $\langle T \rangle$ for $\delta \approx \pi$ is consistent with that of $\langle R/T \rangle$ shown in Fig. \ref{land_res_reg_Ia}:
the transmission reduction at $\delta = \pi$ is in agreement with the resistance enhancement in Fig. \ref{land_res_reg_Ia}.
The physical interpretation of this result will be discussed in the next section.


\section{Discussion of the behavior for $\delta \approx \pi$}
\label{discussion}

The aim of this section is to give a more qualitative and physical explanation of the {\em reversal in the trend of the average resistance and transmission coefficient} as 
$\delta$ approaches $\pi$, a phenomenon which has been described exactly by our mathematical recursion relation.

In Sec. \ref{pert. theory} we found that a perturbative approximation
in the small parameter $y_0$ can be written down analytically and thus
gives a more qualitative description than just the exact numerical solution;
indeed, we were able to describe, within this approximate method, the oscillations as a function of $n$.

We can also employ a similar perturbative approach to describe the average resistance as a function of $\delta$, and investigate whether we can find an indication of the reversal in the trend as $\delta$ moves towards $\pi$.
Of course, we cannot rely on perturbation theory if $\delta$ gets too close to 
$\pi$.
\begin{figure}[h]
\epsfig{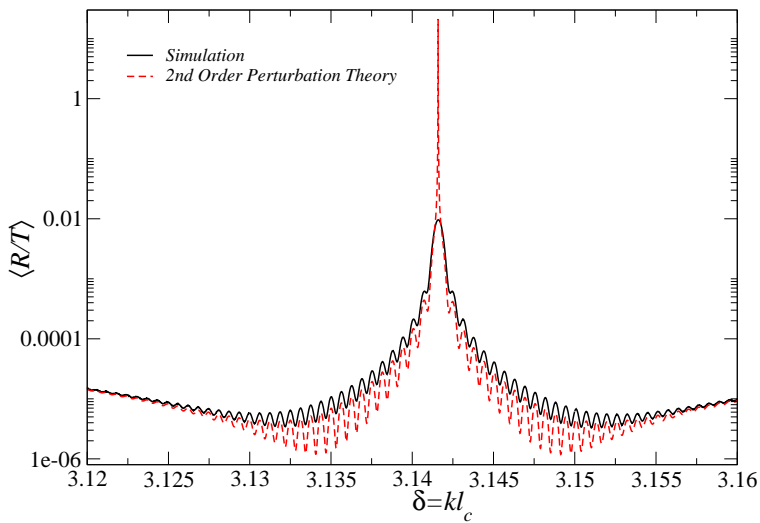}
\caption{
Results of perturbation theory, up to second order for eigenvalues and eigenvectors, and computer simulations for the average Landauer resistance as a function of 
$\delta$, for $n=5000$ scatterers.
The matrix elements of $\Delta \Omega$ of the perturbation were calculated numerically.
The tendency of $\langle R/T \rangle$ to decrease with increasing $\delta$ and subsequently recover as $\delta$ approaches $\pi$ is reproduced by the perturbative approach.
Very close to $\delta = \pi$, the approximation clearly fails.
}
\label{PT_and_sim_resist_vs_delta}
\end{figure}


Figure \ref{PT_and_sim_resist_vs_delta} compares computer simulations of the average Landauer resistance with the results of second-order perturbation theory, as a function of the phase parameter $\delta$ and for a fixed number of scatterers, $n=5000$.
We observe that the tendency of the average resistance to decrease with increasing 
$\delta$ and subsequently recover as $\delta$ approaches $\pi$ is reproduced by the approximate, perturbative approach. 

The behavior of the system that we have described in the above paragraphs is reminiscent of the incipient formation of a band gap that occurs in a finite stretch of an otherwise infinite, periodic Kronig-Penney model.
We now exhibit the similarity of this phenomenon in the the two problems.

\begin{figure}[h]
\epsfig{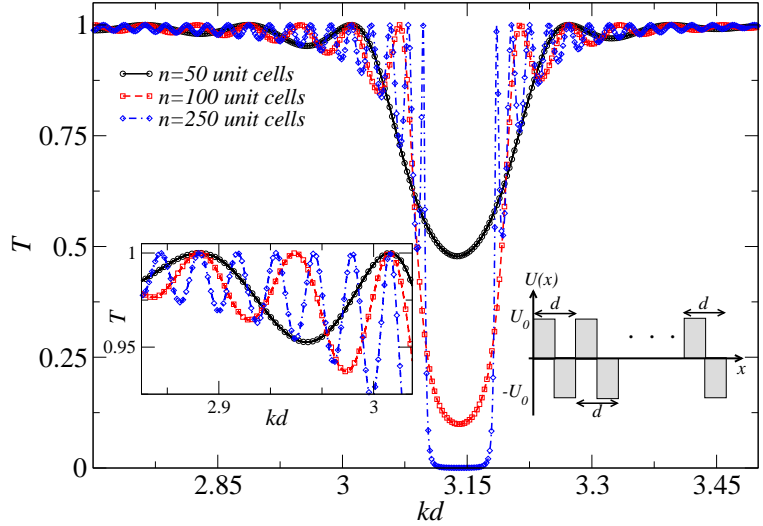}
\caption{
The transmission coefficient for a stretch of 5000 identical scatterers, each consisting of a barrier and a well.
Each unit cell has a length $d$.
In the vicinity of $kd=\pi$, $T$ shows a dip, which becomes ever deeper as $n$ increases.
One observes interference fringes on each side of the dip.
}
\label{T for finite stretch KP}
\end{figure}

The finite stretch of the periodic problem can be formulated by means of a recursion relation in terms of the $2 \times 2$ transfer matrix for the unit cell, assumed to have a length $d$ 
(see inset in Fig. \ref{T for finite stretch KP}), 
as indicated in Eqs. (\ref{M(n+m)=M(m)M(n) KP})
(see, e.g., ref. \cite{merzbacher}, p. 100).
Alternatively, the problem can be also formulated by means of a recursion relation in terms of a $3 \times 3$ matrix, again defined for the unit cell, whose structure is similar to that appearing in Eqs. (\ref{z(n+1)=Mz(n) 2}),
(\ref{z(n),z(0)}) and (\ref{z(n+1)=Mz(n) 1}) for the disordered problem.
With the definitions
\begin{subequations}
\begin{eqnarray}
A(n)
&=& 1+2\vert\beta^{(n)}\vert^{2},
\label{A(n) KP} \\
b(n)
&=& e^{2inkd} (\alpha^{(n)} \beta^{(n)}),
\label{b(n) KP} \\
z(n)&=& \left[A(n)/2, \; (b(n)\sqrt{2})e^{-ikd}, \; (b^{*}(n)\sqrt{2})e^{-ikd}\right]^T \; ,
\end{eqnarray}
\label{A(n),b(n) KP}
\end{subequations}
we rewrite the recursion relations Eqs. (\ref{M(n+m)=M(m)M(n) KP}) as
\begin{equation}
z(n+1) = \Omega_{y_0}^{KP}(kd) z(n) \; ,
\label{z(n+1)=Omega z(n) KP}
\end{equation}
which leads to
\begin{subequations}
\begin{eqnarray}
z(n) &=& (\Omega^{KP})^n z(0) 
\label{sol. z(n) KP} \\
z(0)&=&[1/2,0,0]^T \;.
\label{z(0) KP}
\end{eqnarray}
\label{z(n),z(0) KP}
\end{subequations}
We thus see that in the ordered problem, 
{\em the quantity $kd$} ($d$ being the size of the unit cell) 
which appears in the recursion relations 
(\ref{A(n),b(n) KP}) to (\ref{z(n),z(0) KP}), 
{\em plays a role similar to $kl_c$} for the disordered problem 
($l_c$ being, in this case, the ``minimum unit cell"), 
which enters the recursion relations 
(\ref{z(n+1)=Mz(n) 2}) and (\ref{z(n),z(0)}).

We give evidence for the similarity in the response of the two problems by comparing Fig. \ref{T for finite stretch KP} with Fig. \ref{<T>_vs_delta}.
In Fig. \ref{T for finite stretch KP} we observe the 
{\em incipient formation of a forbidden band}, which manifests itself as a dip in the transmission coefficient $T$ in the vicinity of $kd=\pi$, with interference fringes on each side.
We call it ``incipient", because $n$ is finite.
This is similar to what we observe in Fig. \ref{<T>_vs_delta} for the disordered case, in the vicinity of $\delta=kl_c=\pi$.
In both cases, i) the dip becomes ever more conspicuous as $n$ increases.
This is shown in Fig. \ref{T for finite stretch KP} for the ordered case and was verified for the disordered one.
As a result, in a scattering experiment carried out in this region, the transmission coefficient in the ordered case, and the average transmission in the disordered one, suffer a reduction, with a peak to valley ratio that increases with $n$.
Also, in both cases, 
ii) the dip becomes wider as the strength of the potential increases 
(this we verified by changing $y_0$),
iii) we see interference fringes at the edges, as seen in the insets of 
Figs. \ref{<T>_vs_delta} and \ref{T for finite stretch KP}.
The above behavior is consistent with the one observed for the average resistance, $\langle R/T \rangle$, described at the beginning of the present section.

In the ordered case, the effect discussed above results from the coherent contribution of all the barriers and wells;
indeed, it has been described as the 
{\em collective behavior of the poles of the $S$ matrix} for this problem \cite{gaston}.
In the disordered case, we believe it to be a consequence of the barriers and wells having the same width $l_c$,
and we {\em conjecture} a similar collective behavior.

\section{Conclusions}
\label{concl}

To summarize, we have discussed the problem of wave transport in 1D disordered systems consisting of $n$ weak barriers and wells having a finite, constant width $l_c$, and random strength.  For the calculation of the average Landauer resistance, the problem is reduced to the diagonalization of a three-dimensional complex symmetric matrix.
Approximate results can be obtained analytically, by truncating the matrix when the phase parameter $\delta=kl_c$ is very far from $\pi$ (regime I).
In regime II, the method is improved by using perturbation theory when $\delta$ is not too close to $\pi$. 
When $\delta \approx \pi$ (well inside regime II), the diagonalization was done numerically, giving essentially exact results. 
The average conductance was calculated approximately, making use of Melnikov's equation in regime I and, in regime II, using the results obtained for the resistance.
The theoretical results were verified in the two regimes using computer simulations.

In regime I, the average Landauer resistance was found, for a fixed $\delta$, to increase exponentially with $n$. 
The mfp depends on $\delta$: as $\delta$ increases towards $\pi$, both the  
average Landauer resistance and the average conductance show that the system becomes more delocalized.

As we enter regime II, a new feature appears, compared with older calculations: the transport properties show an oscillatory behavior as functions of $n$ and/or $\delta$, which we could explain using perturbation theory.

Well inside regime II ($\delta \approx \pi$), a second phenomenon shows up: we found an incipient band gap, or forbidden region, where 
i) the average conductance suffers a reduction, and
ii) the average Landauer resistance increases by various orders of magnitude.
In this region, a small change in $\delta$ modifies drastically the transport behavior as a function of $n$.
A more qualitative and physical explanation of this behavior is presented in 
Sec. \ref{discussion}, 
i) in terms of an approximate, perturbative approach, and 
ii) as a reminiscence of the incipient formation of a band gap in a finite stretch or an otherwise infinite, Kronig-Penney problem.

The phenomena we described in the paper and the success of our theoretical analysis in their description suggest the importance of the system's experimental realization.
One possibility we may suggest is in the microwave domain
(see, e. g. Refs. \cite{azi}).
One could construct a medium consisting of plastic pieces, all of the same thickness, but with different indeces of refraction.
One could then shuffle the plastic pieces and create a different random realization of the sample.
The quantity to be measured is the transmission coefficient of each sample.
Another possibility is in the domain of elastic waves in metallic bars.
This is a problem which, in the last years, has received great attention
(see, e.g., Refs. \cite{flores}).
One could construct a bar with indentations and bulges, all of the same length, but with different, random, depths and heights.
A collection of such bars would then constitute an approximation to the ensemble
we need.

\acknowledgments
MY is grateful to the IFUNAM for its hospitality during the development of this work.
PAM acknowledges support from Conacyt, under Contract 79501, and from DGAPA, under Contract PAPIIT IN109014.
The authors are greatful to G. Garc\'ia-Calder\'on for suggesting the analogy with a finite stretch of a Kronig-Penney model.

\appendix

\section{The recursion relation for the average Landauer resistance}
\label{recursion}

From the combination rule given in Eq. (\ref{M(n+m)=M(m)M(n)}) we find a recursion relation for Landauer's resistance of the chain, averaged over the ensemble, as
\begin{subequations}
\begin{eqnarray}
&&\hspace{-1.0cm}
\Big[1+2\langle\vert\beta^{(n+1)}\vert^{2}\rangle\Big]
-\Big[1+2\langle\vert\beta^{(n)}\vert^{2}\rangle\Big]
=2\langle\vert\beta_{n+1}\vert^{2}\rangle\Big[1+2\langle\vert\beta^{(n)}\vert^{2}\rangle\Big]  
\nonumber \\
&&\hspace{6.2cm}
+2\Big[\langle\alpha_{n+1}\beta^{*}_{n+1}\rangle\langle\alpha^{(n)}\beta^{(n)}\rangle + 
{\rm c.c.}\Big] \; ,
\label{rec.rel. 1}  \\
\nonumber \\
&&\hspace{0.55cm}
\langle\alpha^{(n+1)}\beta^{(n+1)}\rangle-\langle\alpha^{(n)}\beta^{(n)}\rangle=
\langle\alpha_{n+1}\beta_{n+1}\rangle\Big[1+2\langle\vert\beta^{(n)}\vert^{2}\rangle\Big]
\nonumber \\
\nonumber &&\hspace{5.7cm}
+\left(\langle\alpha^{2}_{n+1}\rangle -1\right)\langle\alpha^{(n)}\beta^{(n)}\rangle +\langle\beta^{2}_{n+1}\rangle\langle\alpha^{(n)}\beta^{(n)}\rangle^{*},
\\
\label{rec.rel. 2}
\end{eqnarray}
\label{rec.rel. 1,2}
\end{subequations}
where ${\rm c.c.}$ stands for ``complex conjugate".
Using the definitions of Eqs. (\ref{A(n),b(n)}), Eqs. (\ref{rec.rel. 1,2}) take the form
\begin{eqnarray}
\left[
\begin{array}{c}
\frac{A(n+1)}{2} \\
\frac{ib(n+1)}{\sqrt{2}} \\
-\frac{ib^{*}(n+1)}{\sqrt{2}}
\end{array}
\right]
&=&\left[
\begin{array}{ccc}
1 + 2\langle |\tilde{\beta}_{1} |^2 \rangle  
& \sqrt{2}e^{i\delta}\langle \tilde {\alpha}_{1} \tilde {\beta}_{1}\rangle 
&  \sqrt{2}e^{-i\delta}\langle \tilde {\alpha}_{1} \tilde {\beta}_{1}\rangle^{*}  \\
\sqrt{2}e^{i\delta}\langle \tilde {\alpha}_{1} \tilde {\beta}_{1}\rangle  
& e^{2i\delta}\langle \tilde {\alpha}_{1}^2 \rangle
& \langle \tilde {\beta}_{1}^2 \rangle  \\
\sqrt{2}e^{-i\delta}\langle \tilde {\alpha}_{1} \tilde {\beta}_{1}\rangle^{*}  
& \langle \tilde {\beta}_{1}^2 \rangle 
&  e^{-2i\delta}\langle \tilde {\alpha}_{1}^2 \rangle^{*}
\end{array}
\right] 
\left[
\begin{array}{c}
\frac{A(n)}{2} \\
\frac{ib(n)}{\sqrt{2}} \\
-\frac{ib^{*}(n)}{\sqrt{2}}
\end{array}
\right] ,  
\label{z(n+1)=Mz(n) 1}  
\end{eqnarray}
which can be written in the abbreviated form of Eq. (\ref{z(n+1)=Mz(n) 2}).
The $3 \times 3$ matrix appearing in Eq. (\ref{z(n+1)=Mz(n) 1}) will be designated as $\Omega_{y_0}(\delta)$.
It is often useful to write this matrix as 
\begin{subequations}
\begin{eqnarray}
\Omega_{y_0}(\delta) &=& \Omega_0 (\delta) + \Delta \Omega_{y_0}(\delta) \; ,
\label{M=M0+DeltaM a} \\
\Omega_0 (\delta) &=& 
\left[
\begin{array}{ccc}
\mu_1^{(0)} & 0 & 0 \\
0 & \mu_2^{(0)} & 0 \\
0 & 0 & \mu_3^{(0)}
\end{array}
\right]
=\left[
\begin{array}{ccc}
1 & 0 & 0 \\
0 & e^{2i\delta} & 0 \\
0 & 0 & e^{-2i\delta}
\end{array}
\right] \; ,
\label{M0}
\end{eqnarray}
\label{M=M0+DeltaM}
\end{subequations}
the unperturbed matrix $\Omega_0(\delta)$ being the limiting value of 
$\Omega_{y_0}(\delta)$ in the absence of a potential, i.e., for $y_0=0$.

For $\delta = \pi$, the three $\mu_a^{(0)}$ are degenerate and equal 
to 1.
In this case, and for weak scattering, i.e., $y_0 \ll 1$, $\Omega$ takes the approximate form 
\begin{subequations}
\begin{eqnarray}
\Omega_{y_0}(\pi) &\approx&I + y_0^2 \Omega_{red}  
\label{Omega y0<<1}  \\
\Omega_{red} &=& 
\frac{1}{12\pi^3}
\left[
\begin{array}{ccc}
0 & -\sqrt{2} & -\sqrt{2} \\
-\sqrt{2} & -(2\pi + i) & 0 \\
-\sqrt{2} & 0 & -(2\pi - i)
\end{array}
\right] \; ,
\label{Omega-red y0<1}
\end{eqnarray}
\label{Omega,Omega-red y0<<1}
\end{subequations}
where $\Omega_{red}$ is approximately (i.e., for $y_0 \ll 1$) independent of $y_0$.
In the present case, $\Delta \Omega_{y_0}(\delta)$ of Eq. (\ref{M=M0+DeltaM a})
is $\Delta \Omega_{y_0}(\delta) = y_0^2 \Omega_{red}$.

\section{Diagonalization of the matrix $\Omega$, Eq. (\ref{z(n+1)=Mz(n) 1}).}
\label{diagonalizing Omega}

The matrix $\Omega$ is complex symmetric;
provided it has no double characteristic values, it can be diagonalized by a 
{\em complex orthogonal transformation}:
calling 
\begin{equation}
D
=\left[
\begin{array}{ccc}
\mu_1 & 0 & 0 \\
0 & \mu_2 & 0 \\
0 & 0 & \mu_3
\end{array}
\right]
\label{mu's}
\end{equation}
the matrix of eigenvalues and
$O$ the complex orthogonal matrix whose columns are the eigenvectors of $\Omega$, we have
\begin{equation}
\Omega = ODO^T \; .
\label{e-value eqn}
\end{equation}
The new vector
\begin{equation}
z'(n) = O^T z(n)
\label{z'=OTz} 
\end{equation}
has the particularly simple solution
\begin{subequations}
\begin{eqnarray}
z'(n)=D^n z'(0) ,
\label{sol. z'(n)}
\end{eqnarray}
with components
\begin{eqnarray}
z'_a(n)=(\mu_a)^n z'_a(0) .
\label{z'a(n)}
\end{eqnarray}
\label{z'}
\end{subequations}
The original vector $z(n)$ can thus be expressed as
\begin{equation}
z(n) = (O D^n O^T) z(0) \; .
\label{z'=OTz} 
\end{equation}
The first component of this equation gives $A(n)/2$.
Using the initial condition (\ref{z(0)}) and Eq. (\ref{z'=OTz}), we thus find
(assuming that $\Omega$ has no double characteristic values)
\begin{equation}
A(n) = \sum_{a=1}^3 (O_{1a})^2 (\mu_a)^n \; .
\label{A(n) 1} 
\end{equation}
We notice that only the first component of each of the three eigenvectors enters the expression for $A(n)$.

\section{Perturbation theory}
\label{pert. theo.}

We consider the eigenvalue equation
\begin{equation}
\Omega {\bf v}_i = \mu_i {\bf v}_i \; , 
\hspace{1cm} i=1,2,3.
\label{e-value eqn 1}
\end{equation}
The eigenvectors ${\bf v}_i$ were previously designated as the columns of the matrix $O$ of Eq. (\ref{e-value eqn}).
The quantity $A(n)$ of Eq. (\ref{A(n) 1}) can be written in terms of the above eigenvectors as
\begin{equation}
A(n)
= \mu_1^n ({\bf v}_1)_1^2 + \mu_2^n ({\bf v}_2)_1^2 + \mu_3^n ({\bf v}_3)_1^2 \; ,
\label{A(n) 3}
\end{equation}
where $({\bf v}_i)_1$ designates component 1 of the eigenvector ${\bf v}_i$.

If we express $A(n)$ of Eq. (\ref{A(n) 3}) as
\begin{subequations}
\begin{eqnarray}
A(n)&=& \sum_{a=1}^3 A^{(a)}(n) 
\label{A(n) 4 a} \\
A^{(a)}(n) 
&=& ({\bf v}_a)_1^2 \; \mu_a^n \; ,
\label{Aa(n)}
\end{eqnarray}
\label{A(n) 4}
\end{subequations}
we can write
\begin{equation}
\log A^{(a)}(n)
= \log [({\bf v}_a)_1^2]
+ n \log \mu_a.
\label{lnAa(n)}
\end{equation}
The first term in Eq. (\ref{lnAa(n)}) and the coefficient of $n$
are the two parameters of a straight line representing $\log A^{(a)}(n)$ as a function of $n$.
If we develop perturbation theory in the eigenvalues and eigenvectors of $\Omega$ so as to give corrections of the same order in $\Delta \Omega$ in both terms of Eq. (\ref{lnAa(n)}), we shall be building a 
{\em consistent approximation to the two parameters that define the straight line} that we have just described. 
A perturbation theory with this criterion is briefly developed in what follows and used in the main text.
The theory is taken over, almost {\em verbatim}, from the perturbation theory developed in any textbook on Quantum Mechanics, being careful to consider $\Omega$ not as a Hermitean matrix, but as a complex-symmetric matrix.

If we write for the eigenvalues $\mu_i$ of Eq. (\ref{e-value eqn 1}) the expansion
\begin{equation}
\mu_i = \mu_i^{(0)} + \mu_i^{(1)} + \mu_i^{(2)} + \cdots \; ,
\label{e-values expansion}
\end{equation}
we find
\begin{subequations}
\begin{eqnarray}
\mu_i^{(0)}  
&=& \left\{
\begin{array}{cc}
1 \; ,  & i=1    \\
{\rm e}^{2i\delta} \; , & i=2 \\
{\rm e}^{-2i\delta} \; , & i=3
\end{array} 
\right.
\label{mu(0)}   \\
\mu_i^{(1)}  
&=& \Delta \Omega _{ii}         
\label{mu(1)}   \\
\mu_i^{(2)}
&=& \sum_{j (\neq i)} 
\frac{\Delta \Omega _{ij} \Delta \Omega _{ji}}{\mu_i^{(0)}- \mu_j^{(0)}} 
\label{mu(2)}  \\
\cdots \nonumber
\end{eqnarray}
\label{pert exp mu}
\end{subequations}
Similarly, for the eigenvectors ${\bf v}_i$ of $\Omega$ we write the expansion
\begin{equation}
{\bf v}_i = {\bf v}_i^{(0)} + {\bf v}_i^{(1)} + {\bf v}_i^{(2)} + \cdots \; ,
\label{e-values expansion}
\end{equation}
and find
\begin{subequations}
\begin{eqnarray}
{\bf v}_i^{(0)}  
&=& \left\{
\begin{array}{cc}
(1, 0, 0)^T \; ,  & i=1    \\
(0, 1, 0)^T \; , & i=2 \\
(0, 0, 1)^T \; , & i=3
\end{array} 
\right.
\label{v(0)}   \\
{\bf v}_i^{(1)}  
&=& \sum_{j (\neq i)} \frac{\Delta \Omega _{ji}}{\mu_i^{(0)}- \mu_j^{(0)}} 
{\bf v}_j^{(0)}
\label{v(1)}   \\
{\bf v}_i^{(2)}
&=& \sum_{j,k (\neq i)} 
\frac{\Delta \Omega _{jk} \Delta \Omega _{ki}}
{(\mu_i^{(0)}- \mu_j^{(0)})(\mu_i^{(0)}- \mu_k^{(0)})} 
{\bf v}_j^{(0)} 
\nonumber \\
&&-\sum_{j (\neq i)} 
\frac{\Delta \Omega _{ii} \Delta \Omega _{ji}}
{(\mu_i^{(0)}- \mu_j^{(0)})^2} 
{\bf v}_j^{(0)} 
-\frac12 
\left[ \sum_{j(\neq i)} 
\frac{\Delta \Omega _{ij} \Delta \Omega _{ji}}
{(\mu_i^{(0)}- \mu_j^{(0)})^2} \right]
{\bf v}_i^{(0)} 
\label{v(2)}  \\
\cdots \nonumber
\end{eqnarray}
\label{pert exp v}
\end{subequations}

Substituting these results in Eq. (\ref{A(n) 3}), we can verify the identity $A(0)=1$ up to second order in $\Delta \Omega$.

\section{Recursion relations for a finite stretch of a periodic Kronig Penney model}
\label{recursion KP}

A finite stretch of a Kronig-Penney problem obeys the recursion relation 
\begin{subequations}
\begin{eqnarray}
\textbf{\textit{M}}^{(n+1)} 
&=&\textbf{\textit{M}}_{n+1} \textbf{\textit{M}}^{(n)}   \\
&=& D^{-1}((n+1)kd) \textbf{\textit{P}}^{n+1} \; , \\
{\rm where}\;\;\;\;\; \textbf{\textit{P}} &=& D(kd) \mathaccent 23 {M}_{1}  \\
{\rm and} \;\;\;\;\; 
D(kd) &=&
\left[
\begin{array}{cc}
e^{ikd} & 0 \\
0 & e^{-ikd}
\end{array}
\right] \; ,
\end{eqnarray}
\label{M(n+m)=M(m)M(n) KP}
\end{subequations}
written in terms of the $2 \times 2$ transfer matrix for the unit cell, assumed to have a length $d$ (see, e.g., ref. \cite{merzbacher}, p. 100);
here, $\mathaccent 23 {M}_{1}$ is the transfer matrix for the unit cell translated to the vicinity of the origin.

Alternatively, with the definitions (\ref{A(n),b(n) KP}),
we write Eqs. (\ref{M(n+m)=M(m)M(n) KP}) as (\ref{z(n+1)=Omega z(n) KP}),
where 
\begin{eqnarray}
\Omega_{y_0}^{KP}
&=& \left[
\begin{array}{ccc}
1 + 2 |\mathaccent 23 {\beta}_{1} |^2   
& \sqrt{2}e^{ikd} (\mathaccent 23 {\alpha}_{1} \mathaccent 23 {\beta}_{1}^{*}) 
&  \sqrt{2}e^{-ikd} (\mathaccent 23 {\alpha}_{1} 
\mathaccent 23 {\beta}_{1}^{*})^{*}  \\
\sqrt{2}e^{ikd} (\mathaccent 23 {\alpha}_{1} \mathaccent 23 {\beta}_{1}) 
& e^{2ikd} \mathaccent 23 {\alpha}_{1}^2 
& \mathaccent 23 {\beta}_{1}^2   \\
\sqrt{2}e^{-ikd} (\mathaccent 23 {\alpha}_{1} \mathaccent 23 {\beta}_{1})^{*}  
&  (\mathaccent 23 {\beta}_{1}^*)^2 
&  e^{-2i kd} (\mathaccent 23 {\alpha}_{1}^*)^{2}
\end{array}
\right] \; .
\label{Omega KP}
\end{eqnarray}
This leads to Eqs. (\ref{z(n),z(0) KP}).

\section{Reduction to the results of the dense weak-scattering limit}
\label{steps_vs_dwsl}

In this appendix we briefly investigate the limit in which the results of the present model --consisting of finite-size scatterers-- reduce to those obtained in the dense weak-scattering limit (DWSL) of 
Ref. \cite{froufe_et_al_2007}, consisting of a succession of delta scatterers.

\subsection{The present model}
\label{present model}

A barrier lower than the energy requires (see Eq. (\ref{sogamoso}))
$y_r < \delta^2$, so that very weak barriers are characterized by 
$y_0 \ll \delta^2$. 
We further require the wavelength $\lambda$ to be much larger than the barrier 
width $l_c$, i.e., $\delta = k l_c \ll 1$. 
We thus have the joint requirements
\begin{equation}
y_0 \ll \delta^2 \ll 1 \; .
\label{weak+long_wl 1}
\end{equation}

Eq. (\ref{<beta1^2> 2}) for the mfp (designated here by $\ell$) can be written in the equivalent ways
\begin{subequations}
\begin{eqnarray}
\frac{1}{k \ell} &=& \frac{y_0^2}{12 \delta^3}  \; , 
\label{1/kl}     \\
\eta &\equiv& \frac{1}{\nu \ell}
=\frac{y_0^2}{12 \delta^2} \; ,
\label{eta}     \\
\delta &=& \frac{y_0}{\sqrt{12 \eta}} \; ,
\label{delta(y0,eta)} \\
\frac{1}{k \ell} &=& \frac{\eta}{\delta} \; ,
\label{1/kl = eta/delta}
\end{eqnarray}
\label{1/kl,eta,delta}
\end{subequations}
$\nu = 1/l_c$ being the density of scatterers.
A problem is thus specified by the three parameters $\eta, \; y_0, \; \delta$, 
related by one of the above equations, like (\ref{delta(y0,eta)}).
To satisfy the inequality (\ref{weak+long_wl 1}) we need
\begin{equation}
12 \eta \ll y_0 \ll \sqrt{12 \eta} \; .
\label{weak+long_wl 2}
\end{equation}
We follow the steps:

i) propose $\eta \ll 1$;

ii) propose $y_0$ to be consistent with (\ref{weak+long_wl 2}); this is used to set up the numerical barrier model.

iii) find $\delta$ from (\ref{delta(y0,eta)}).

\subsection{The DWSL model}
\label{dwsl}

The DWSL model of Ref. \cite{froufe_et_al_2007} consists of a succession of equally spaced (spacing = $d$) delta potentials, with an rms intensity $u_0$, having units of $k$. 

The relation defining the mfp can also be written in various equivalent ways
\begin{subequations}
\begin{eqnarray}
\frac{1}{k \ell} 
&=& \frac{u_0^2}{12 k^3 d} \; ,
\label{1/kl-dwsl a}  \\
&=& \frac{v_0^2}{3 k d} \; ,
\label{1/kl-dwsl b}  \\
\frac{d}{\ell}&=&\frac{v_0^2}{3} \; .
\label{d/l-dwsl}
\end{eqnarray}
\end{subequations}
Here, $d$ is the distance between successive delta potentials and $v_0 = u_0/2k$.

In this model, too, the problem is specified by three parameters:
$k \ell, \; kd, \; v_0$, related by one of the above equations, like 
(\ref{d/l-dwsl}).

\subsection{Connection between the two models}
\label{dwsl}

We need to connect the two models:

i) choose $k \ell$ to be the same in the two models

ii) choose $kd$ of the DWSL delta-potential model to coincide with $kl_c = \delta$ of the finite-size scatterer model. 
This implies that the fraction of wavelength contained in the interval between 
the centroids of two successive scatterers is the same in the two models
(compare Fig. \ref{rand_steps} of the present paper with Fig. 3 of Ref. \cite{froufe_et_al_2007}).

iii) from $kd$ and $k \ell$ we find $d/ \ell$ and hence $v_0$ from 
Eq. (\ref{d/l-dwsl}), which is to be used to set up the numerical delta-potential model.
\begin{figure}[h]
\epsfig{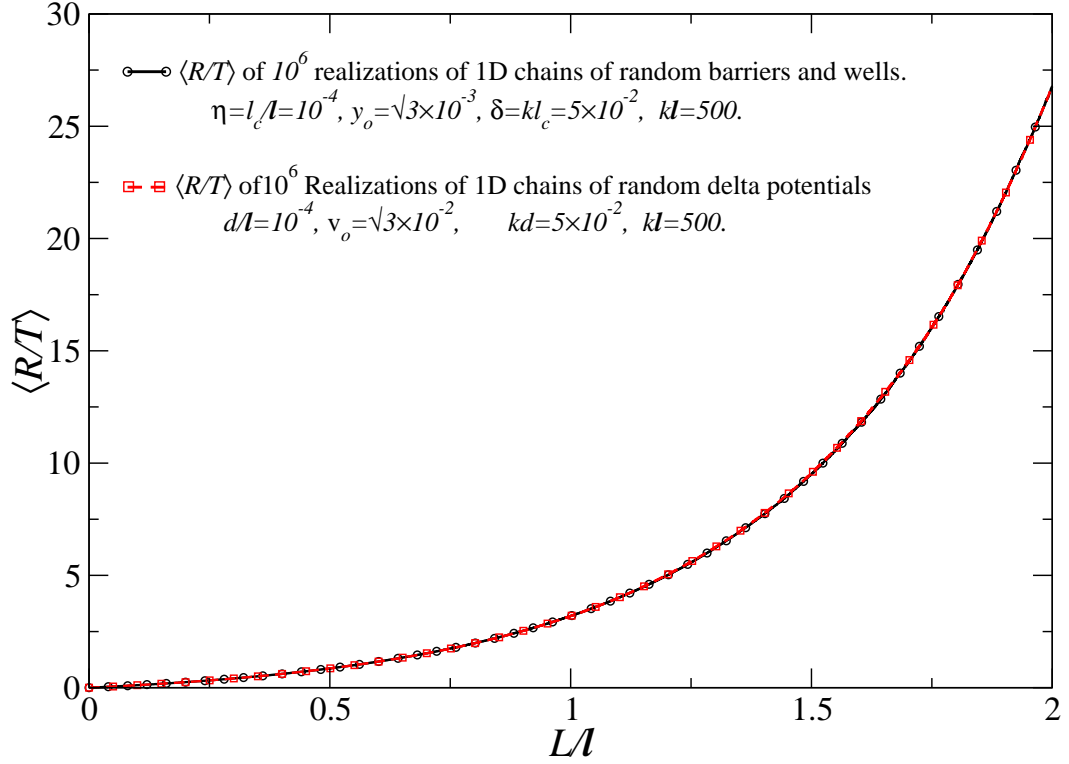}
\caption{
Results of computer simulations for the average Landauer resistance for the present model and the DWSL model of Ref. \cite{froufe_et_al_2007}.
The parameters chosen for each model are indicated in the figure, and conform to the criteria explained in the text. 
The agreement is excellent.
}
\label{steps_vs_dwsl}
\end{figure}

Fig. \ref{steps_vs_dwsl} shows computer simulations for the average Landauer resistance for the two models as a function of $L/\ell$, $L$ being the length of the chain, for the parameters indicated in the figure. 
The agreement is excellent.
This figure is similar to Fig. 3 of Ref. \cite{froufe_et_al_2007}).

\clearpage


\end{document}